# Automatic Feature Extraction, Categorization and Detection of Malicious Code in Android Applications

Muhammad Zuhair Qadir
Department of Computer Sciences
School of Science and Engineering
Lahore University of Management and Sciences
Lahore, Pakistan
mzuhairqadir@gmail.com

Atif Nisar Jilani
Department of Computer Sciences
School of Science and Engineering
Lahore University of Management and Sciences
Lahore, Pakistan
jilani.ansar@gmail.com

Hassam Ullah Sheikh
School of Computer Science
University of Manchester
Manchester, UK
sheikhh@cs.man.ac.uk

*Abstract*—Since Android has become a popular software platform for mobile devices recently; they offer almost the same functionality as personal computers. Malwares have also become a big concern. As the number of new Android applications tends to be rapidly increased in the near future, there is a need for automatic malware detection quickly and efficiently. In this paper, we define a simple static analysis approach to first extract the features of the android application based on intents and categories the application into a known major category and later on mapping it with the permissions requested by the application and also comparing it with the most obvious intents of category. As a result, getting to know which apps are using features which they are not supposed to use or they don't need.

## I. Introduction

Android has seen a tremendous growth since its introduction in the market. The numbers of applications are growing with extraordinary volume. It has a wide of variety of applications targeting various needs and use cases. Since its launch the privilege to the developers who can upload any application draws attention to the malicious developers whose intents are to deceive and take advantage of this privilege and without much inspection it is getting hard to investigate a valid and benign application which can filter out the malicious application from the pool of millions of apps. Malicious code is being injected into mobile applications and threatens the privacy of user's personal data and device integrity also it can lead to breaches of user data and violate application security policies. In 2011, malware attacks are increased by 155 percent across all platforms [1] in particular, Android is the platform with the highest malware growth rate by the end of 2011.

Static detection techniques (also called signature matching) have high detection rates and consume fewer resources [8]. Static analysis of these applications can result in faster detection of malicious apps and as it involves automatic application code lookup and detecting required content without running it or testing it. Static analysis of Android applications is important because quality and reliability are keys to success on the Android market [2].

Because of diversity of applications available in different markets, there is a need to categories the application first.

We perform static analysis on Android applications in order to determine the feature set of the application based on its functionality. Our program also looks for the category in which the app should fall.

## II. Related Work

There is lot of study on dynamic and static approaches for android malware detection analysis. Static being faster as it does not require runtime cost and exploiting the code on runtime.

Confidentiality and authorization are two key goals which are addressed in program analysis techniques. SCanDroid extracts security specifications from the manifest of an app and checks whether data flows through the app are consistent with the stated specifications [3].

Tabe1 [4] gives the crisp comparison of the tools being made to address various malicious code detection and analysis with various techniques.

They are various tools written on static analysis among them the most famous one used in AndroGuard [5] it's an open source project to statically detect Android malwares, it basically reverse engineer APK from byte code (assembly source code) to readable format and afterward it visualizes your application with Gephi which is a control flow graph of the method. Check if an android application is present in a database or not, it maintains the list of pre-defined malware's. Also it compares the control flow graph of both the application to check the similarity between that malware and the given application.

There are various papers which works on the permission analysis approach, one [7] of them detects whether an app is over privileged or underprivileged/ This can be very useful as developers develop apps and without much concern adds more permissions to the app whereas the code itself requires very less number of permissions comparatively.

| Solution | Aim | Flow Analysis | Classification Policy | Evaluation Scale |
|---|---|---|---|---|
| SCanDroid | Enforcement of confidentiality, integrity | Data, string | Constraints on permission logics | N/A |
| CHEX | Discovery of exposed component API | Data | Component exported to public without restrictions | 5,486 apps |
| RiskRanker | Detection of abnormal code/behavior patterns | Data, control | Multiple malware behavior signatures | 118, 318 apps |
| Woodpecker | Firmware permission | Data, control | N/A | 8 phone images, 13 permissions |
| AndroidLeaks | Confidentiality | Data | Sensitive data used by risky APIs | 24,350 apps |
| SCANDAL | Confidentiality | Data | Sensitive data used by risky APIs | 90 apps & 8 malware |
| Stowaway | Detection of overprivileged apps | String, Intent control flow | Compare required and requested permissions | 940 apps |
| ComDroid | Detection of apps communication vulnerabilities | Intent control flow | Implicit Intent with weak or no permission | 100 apps |
| PiOS | Confidentiality | Data | Sensitive data used by risky APIs | 1,407 apps |
| UID | Identification of unauthorized calls | Data, event specific control | Trigger-operation dependence for privileged function calls | 708 apps & 482 malware |

TABLE I. Comparison

Crowdroid [12] is a machine learning-based framework that recognizes Trojan-like malware on Android smartphones, by analyzing the number of times each system call has been used by an application during the execution of an action that requires user interaction. A genuine application differs from its trojanized version, since it issues different types and a different number of system calls. Crowdroid builds a vector of m features (the Android system calls).

Another technique [19] which monitors both the smartphone and user's behaviors by observing that continuously monitors various features and events obtained from the mobile device from sensors activities to CPU usage. And then apply various algorithms like mining techniques to classify the collected data as normal or abnormal. The main assumption in this techniques that system metrics such as CPU consumption, number of sent packets through the Wi-Fi, number of running processes, battery level etc. can be employed for detection of previously un-encountered malware by examining similarities with patterns of system metrics induced by known malware.

Another paper [22] proposed a malicious application detection framework in which it uses both static and dynamic detection technique. Uses an automatic feature extraction tool on android market built in Javascript based on permissions, the framework performs a static detection based on methods of System API calls and performs dynamic detection using machine learning on android market.

Basically it does static analysis on android application output of the readelf tool which extracts their system calls and then they are compared with the pre-defined list of malicious applications from benign ones based on the combinations of system calls used in the executable [8].

### III. APPROACH

We propose and implemented an approach to detect malicious applications statically. Android applications can interact with other applications, and with the system, through a well-defined API. A number of components can make up an application. In particular, Android defines activities, services, content providers, and broadcast receivers.

Activities, services, and broadcast receivers are activated by intents, i.e., asynchronous messages exchanged between individual components to request an action. Activity and service intents specify actions to be performed. Conversely, broadcast receiver intents define the received event and are delivered to the interested broadcast receivers. Our algorithm consists of three main steps.

Step I: We get the set of APKs to analyze and transform to decompressed files and then into byte code using APKTool [6].

Step II: Then we extract features of the application present in the byte code.

Step III: Categorize the application to known major categories.

Step IV: Relate features to category and point out features which are not needed for the application and can be considered as malicious.

In this four step process, APKTool does the conversion part to get the readable format code of the original APKs. While in the second step, feature extraction or more appropriately, code tagging is done through our custom tool written in C++ which parse the code and gets all the intents of the code and tags it with the detected features which will be used later.

Intents provide an easy way to detect what an application actually trying to do and how it is utilizing to the resources of the device. This step is crucial and essential part of the algorithm making an intelligent system to separate benign apps and malicious apps. Our tool is flexible as takes a file full of intents which should be extracted and tagged in the code. Making a useful plugin for various alterations of this algorithm and also targeting specific type of malicious codes which exploits certain features only.

| No | List of Some Targeted Intents |
|---|---|
| 1 | android.hardware.Camera.PictureCallback |
| 2 | android.telephony.SmsMessage |
| 3 | android.telephony.SmsManager |
| 4 | android.telephony.CellLocation |
| 5 | android.media.AudioRecord |
| 6 | android.location.LocationManager |

TABLE II. Intents

Categorization can be done on the basis of group of intents and manifest file. Clustering can tell specific APK falls in which android app category. We made another tool in C++ for to achieve this categorization.

IV. IMPLEMENTATION

We used APK tools for extracting code out of APK into smali code [10] which is actually sort of byte code along with Dalvik Opcodes [11] and having its own syntax. APK tools is a very powerful tool available at android's developer website for various apk reverse engineering tasks. We wrote a batch script which takes the application name , and after wards with the help of APK tool, it extracts all the folder that are compressed within APK into Dalvik code which is not human readable for that purpose batch in the next step converts it into samli code which is byte level human readable code.

Custom C++ tool was written to tag features parsing whole bunch of smali files obtained from previous step. In the tool, a file is passed as parameter having all features to be tagged with the code line number and count of that feature in a specific smali byte code file.

For categorization we wrote another custom tool in C++ to categorize the application, we analyze the group of intents and manifest file for permissions to get most of the intents and features of the app helping us to understand the type of the application whether it is a game, utility, image tool, sound recorder etc. An isolated and unprivileged application has very limited functionality. Therefore, smartphone platforms allow access to individual sensitive resources (e.g., address book, GPS) using permissions. Permission is a form of capability. In order to categorize we have maintained a list of major permissions by analyzing various applications before writing our tool. On the basis of these permissions, we compare the permission presented in manifest file with our list, and also we have set of rules that if an application has access to certain set of permissions then we can say that it may belongs to a certain category as we mentioned in our Table III. If the application is from Google's application market i.e. play then we also take into consideration the category [23] assigned in the Google app store.

| No | List of Categories | Set of Some Permissions |
|---|---|---|
| 1 | Communication | android.permission.WRITE_SMS<br>android.permission.SEND_SMS<br>android.permission.CALL_PHONE<br>android.permission.READ_SMS |
| 2 | Games | android.permission.INTERNET<br>android.permission.READ_PHONE_STATE |
| 3 | Social App | android.permission-group.LOCATION<br>android.permission.READ_CONTACTS<br>android.permission.READ_SOCIAL_STREAM<br>android.permission-group.ACCOUNTS<br>android.permission.INTERNET |
| 4 | Utility | android.permission.BATTERY_STATS<br>android.permission-group.SYSTEM_TOOLS<br>android.permission.BLUETOOTH_ADMIN<br>android.permission.KILL_BACKGROUND_PROCESSES |
| 5 | Education | android.permission-group.STORAGE<br>android.permission.READ_EXTERNAL_STORAGE |
| 6 | Media | android.permission.CAMERA<br>android.permission.RECORD_AUDIO<br>android.permission.MODIFY_AUDIO_SETTINGS<br>android.permission.INTERNET |
| 7 | Widgets | android.appwidget.action.APPWIDGET_UPDATE<br>android.appwidget.action.APPWIDGET_CONFIGURE |
| 8 | Travel & Local | android.permission-group.LOCATION<br>android.permission.INTERNET |

TABLE III. Categories

In the last phase, we use tagged code and categorized APK and point out the features malicious for the application and as a result classifying whether the concerned APK should be considered as malicious or not.

## V. Future Work

We had another approach still to implement to improve the results, in which we will use machine learning approach to further classify the detected malicious to pinpoint the category of malware. So two level filtering will further decrease the false positives and give more accurate results.

Basically the behavior of smart phones is rather protected by the use permissions, also there are numerous permission-protected [20] method calls that are not part of the public Android API, but are in classes that are resident on the phone, we will then examines all the obtained smali files to find method calls used by application and each method call is then compared to the list of all method calls that we have in our list of permission-protected Android API calls to build an association. That association set is then compared to the permission set that is declared in the application's AndroidManifest.xml file, in this way we can determine whether the application has extra permissions, lacks permissions, or has exactly the permission set that it requires based on its functionality.

Furthermore, to make more refine categorization of applications we aim to implement the scheme as implemented in LACTA [14]. LACTA finds certain keywords and does code analysis of the application and based on learning certain keywords and function names it categorizes the application. This will add two level filter on categorization step of our algorithm making it fine-tuned and more effective.

## VI. Conclusion

We have devised a simple approach for automatic static analysis which is capable of source code tagging with its prominent features and application categorization which helps in identifying irrelevant features which should not be present in the app. It is quick and efficient and relies on intents present in the source code. It can be made more efficient using machine learning techniques to train on apps first then predicting the malicious code snippets but it has a tradeoff with performance.

We have also proposed additional filters in the future work that could be added to our approach to increase the viability and accuracy of our system.